\documentclass{article}

\usepackage{mathptmx}       
\usepackage{helvet}         
\usepackage{courier}        
\usepackage{type1cm}        
\usepackage{makeidx}         
\usepackage{graphicx}        
\usepackage{multicol}        
\usepackage[bottom]{footmisc}

\usepackage{latexsym}
\usepackage{enumerate}
\usepackage{hyperref}
\usepackage{url}
\newcommand{\COMMENT}[1]{}
\makeindex             

\begin{document}



\title{Taking Physical Infinity Seriously}
\author{Don Perlis \thanks{{My thanks for helpful comments and clarifications from: Paulo Bedaque, Juston Brodie, Jeff Bub, Jean Dickason, 
Sam Gralla, Dan Lathrop, Carlo Rovelli, Ray Sarraga, and two anonymous reviewers  -- none of whom however is 
to be blamed for any errors or outrageousnesses that remain.} 
}
}
%

\maketitle

\begin{abstract}\
  The concept of infinity took centuries to achieve recognized status
  in the field of mathematics, despite the fact that it was implicitly
  present in nearly all mathematical endeavors.  Here I explore the
  idea that a similar development might be warranted in
  physics. Several threads will be speculatively examined, including
  some involving nonstandard analysis.  While there are intriguing
  possibilities, there also are noteworthy difficulties.
\end{abstract}


\setcounter{footnote}{0}

\section{Introduction}
Infinity plays a central role in mathematics, and arguably always has
-- despite occasional negative characterizations (even by some of the
most esteemed practitioners).  Today surely there is little question
about its importance in the minds of the vast majority of
mathematicians.\footnote{In \cite{davisprag} Martin Davis includes a discussion of infinity in mathematics in
  terms of imaginative powers of our minds (my words, not his), and
  (partly) justifies this by analogy with physics -- somewhat the
  reverse of my point here, but one I am equally sympathetic to.  }
There is also very wide appreciation of the idea that whither goes
mathematics, there also goes physics (and often the other way
around). And yet in physics the notion of infinity plays a rather
curious ``fix-it-up'' role, rather like duct tape, that is brought out
for use whenever needed but then put firmly back in the tool box
again. Thus it is not kept front and center in actual physical models,
quite unlike its now central and fundamental role in
mathematics.\footnote {One prominent example that will {\em not} be
  discussed at any length here are the divergent Feynman integrals
  (among others) of quantum field theory (QFT). See for instance the
  excellent Wikipedia entry for Renormalization \cite{wiki:renorm}.}

This is part of a much larger issue: how mathematics relates to
physical reality. This involves many aspects that we will not touch on
here, other than some brief comments. For instance, Wigner
\cite{wigner1960unreasonable} regards it as ``unreasonable'' that
there is such a strong connection between math and physics.  And
Kreisel \cite{kreisel1974notion} has considered whether quantities
that are physically observable (according to a given physical theory)
can be generated by a Turing machine; such a theory he calls
``mechanical''. See also \cite{aguirre2011cosmological},
\cite{misner1981infinity}, \cite{rovelli2011some}, all of whom discuss
cosmological issues such as whether space is infinite in extent;
Rovelli \cite{rovelli2011some} in particular distinguishes -- similarly to a distinction
we shall draw -- between infinite divisibility and infinite extent.

A related question is: what sort of universe is needed in order for
there to be a possibility of mathematics at all? That is, not actual
mathematical practice, but simply the possibility of ``stuff''
sufficient to allow, for instance, such things as sequences, records,
relations. There would seem to be a requisite minimum level of temporality and
spatiality even for natural numbers to have any meaningfulness. And,
perhaps deeper: what counts as stuff, and what is it for stuff to
``be''? But we will leave these questions aside, and return to our
main theme.\footnote 
{I can't resist noting that in roughly 1968-9
  Martin Davis mentioned to me that in his estimation a huge unclarity
  underlay foundational issues in mathematics and in particular set
  theory: what counts as a thing?}

Here I will describe a number of examples in which infinity is used
explicitly in physics, and possible developments that these might
suggest, including a few detours along the way.\footnote {That the
  topic is appropriate to a volume dedicated to Martin Davis, I
  justify with the observations that (i) Martin helped instill in me a
  general love for ideas on topics far and wide; and (ii) at least two
  of Martin's writings bear on related themes: nonstandard analysis
  \cite{davis2014applied} and quantum physics
  \cite{davis1977relativity}. I note that Rovelli \cite{rovelli1996relational}
  entertains an idea already present in \cite{davis1977relativity},
  namely that of observer-dependent reference frames in quantum
  mechanics; and (personal note from Rovelli) this also apparently has
  come up in writings of Kochen and Isham as well, all after Martin's
  contribution appeared. See also \cite{van2010rovelli} for more on this theme.}  
Yet I must
add that, as a non-physicist, I also approach the broader topic with
some trepidation; and while I have consulted a number of physicists in
the writing of this paper, still any misconceptions are completely my
own.  I trust the reader will pardon any sense that I am throwing in
the kitchen sink; this essay represents some possibly far-flung
imaginings that perhaps do not fall altogether within customary styles
in scientific writing.

The rest of this paper is organized as follows: We describe the
examples just referred to above, to distinguish several modes of use of
infinities in physics; next I review some ideas due to Jose Benardete
on a Zeno-like puzzle about infinity, and some related issues
concerning particles, densities, and spin; we then turn to
nonstandard analysis as one methodology that appears to shed some
light (in connection with Dirac delta functions), but has difficulties
of its own.

\section{Multiple uses of infinity in physics}
Quantum mechanics provides us with many intriguing examples of our
subject; I give three here. First, Schr{\"o}dinger's solution of his wave
equation for the energy levels of the hydrogen atom involves an
argument in which infinity plays the role of a kind of {\em reductio},
or proof by contradiction, leading to the rejection of the infinity. 
Second, that same solution results in an
infinite set of energy levels, which 
are pointedly {\em not} rejected. Third, Dirac
introduced the (infinite-valued on an infinitesimal interval)
delta function because it provided a highly simplifying and
intuitively satisfying notation for his vastly influential treatment
of quantum mechanics. I briefly summarize each of these uses of infinity below.

In a 1926 paper, Schr{\"o}dinger solved his famous wave equation for the
special case of the hydrogen atom.  Along the way he had to set to zero
certain series terms, since otherwise they would lead to variables with infinite
values. (The remaining terms provide 
solutions for energy levels of the hydrogen atom that are the familiar Bohr
ones that closely match experiment\footnote
{
E.g., when associated to the spectral lines found by Balmer in 1886.
} -- but not 
quite close enough; later refinements were needed, including spin and
relativistic effects.)  So in this case, a variable taking on an
infinite value is used as a reason to reject it and instead consider only
alternative lines of argument. This of course is not new to
Schr{\"o}dinger but in fact is a common form of argument, applicable whenever the
variable in question is something one has reason to think should be
finite. I provide this particular example of such a {\em reductio} use of
infinity here (as opposed to any number of others) simply because it
is curious that it arises in the same setting in which the next
example occurs.  We may refer to this first as a {\em dense} physical
infinity: a physical variable (that in principle might be measured by
means of instruments within certain physical confines) taking on (but
perhaps should not do so) an infinite value. This is employed via a {\em
  reductio} to eliminate the infinity (sometimes easily as above,
sometimes with enormous effort and controversy as in QFT).

Yet a result of Schr{\"o}dinger's argument is that the
distinct possible energy levels of the hydrogen
atom alluded to above are infinite in number, and in fact a specific formula is derived for the possible
energies, $E_{n}$ where $n = 1,2,...$. {\em This} infinitude is not
shrugged off as unphysical; each and every $E_{n}$ is taken as
representing an in-principle possible physical energy for the atom.
\footnote
{A very recent result \cite{friedmann_hagen_2015} even derives the famous centuries-old 
Wallis formula for $\pi$ from the very same infinite sequence of hydrogen's energy levels, 
something no one had the faintest idea could happen, suggesting that the infinitude has yet 
further significance -- although just what that may be is unclear.
}
Indeed, it is the excellent match-up with experiment that makes the Schr{\"o}dinger 
result so convincing.\footnote {For instance, had Schr{\"o}dinger's
  calculation led instead to a sequence of values for $E_{n}$ that
  stopped after $n=20$, surely there would have been a frenzied
  attempt by experimentalists to find twenty-one energy levels to test
  the theoretical result.}  Of course, it is similar in kind to the
infinitude of possible heights (or potential energies) of a projectile
above ground level, which is also not seen as unusual. These perhaps
amount, in the end, to little more than the fact that the infinite
(unbounded) set of real numbers, {\bf R}, is taken as the possible
range of values for many physical variables (with some limitations as
dictated by a given situation -- but the infinitude is not in general
ruled out). This is a {\em range-of-values} physical infinity: a mere
listing of possible values, of which there may be infinitely many. Yet
it is a possibility that, in some sense, describes (a working picture
of) the universe: the universe has in it an unbounded range of
allowable values for certain variables.\footnote {The chapter by Blass
  and Gurevich in this volume similarly comments on ``infinitely many
  possible values, for example of position or momentum'' and the
  corresponding infinite-dimensional Hilbert space of such a system's
  states. This is closely realted to the idea of an infinite extent of
  space, which may or may not be the case -- but such is not seen as a
  reason to reject a model outright. Similarly, the infinitely-many
  possible reference frames in quantum mechanics suggested in
  \cite{davis1977relativity} is not suspect on the basis of the
  infinity involved.}

One way to make these two standard physical uses of infinity more
intuitive may be this: if a variable represents a measurable quantity,
something that one might detect in an experiment, then the measured
value must be finite: we have no means to measure an actual infinity;
whereas any -- even an infinite -- {\em number} of finite
values might be measured (given enough time). Or: there may be an infinite amount of
space, matter, or energy, in the universe; but not right where the measuring
instruments are located. Note that we are not taking a stand on such a
view; in fact, we are exploring alternative possibilities!

Indeed, one can reason: there may be things physically present that we
cannot measure. One such that comes to mind is the wavefunction
itself; this is sometimes\footnote
{
More so some decades ago; it seems now a minority view.
}
characterized as the fundamental ``reality''
of which our measurements ferret out (and even modify) some features
but never reveal the full thing in itself. If the wavefunction is
really there, yet never fully revealed, why not also infinite energies
and other quantities?  Or consider space and time (or spacetime)
themselves: we never measure all of space or time, by any means. Yet
in measuring bits and pieces, we convince ourselves that there is a
great deal more, and in the case of some theories even that the universe
has an infinitude of such pieces, either extended (range-of-values) or densely packed.

Our third example is Dirac's delta function. This is in wide use by
physicists (and not only in quantum mechanics). Yet the delta function
is routinely viewed as a useful fiction, not something to take
seriously except as a convenient shorthand for a much more cumbersome
and less intuitive set of tools. This mode we then call the {\em useful
fiction} infinity: we use it but we don't believe it corresponds to
anything physical.\footnote
{This is reminiscent of the early uses of imaginary
 numbers: they were clearly useful, but it was far less clear that such
 a number could be a {\em thing} in any sense available back
 then. Eventually two developments helped: (i) the observation that
 imaginary numbers can be interpreted as rotations, and (ii)
 formal/abstract methodology: if something has a consistent
 mathematical use, that is all that is needed in order for it to {\em
 be} an object of mathematical study.}
Nonetheless, it seems to fall also into the {\em dense} mode of
infinity.

Thus we have cases where a dense infinity is outlawed (by reductio),
and others where it is accepted as a useful fiction; and there are
also cases (range-of-values) where infinity is accepted as quite
physically sensible. Much of what we are considering here is whether
some of the ``fiction'' cases should perhaps be considered as less
fiction and more real physics. Delta functions are one case in point
(we shall return to them below) but not the only one.

\section{Benardete's challenge}
Jose Benardete \cite{benardete1964infinity} discusses novel variants of a paradox of
Zeno. Here is a version that suits our purposes: Imagine that an
impenetrable barrier is erected at each point $x=1/2^n$ for n=1,2,...;
we suppose the barriers to be of zero thickness (or of decreasing
thickness as they close in on $x=0$, so that they do not overlap or
touch each other, and so that they do not overlap or touch
$x=0$). Moreover, imagine that each barrier is immovable once so
placed. Finally, imagine that a projectile is aimed at the barriers
from a point to the left, i.e., from some $x<0$. 

Let us first of all note that this appears to be a case of dense
infinity. There is an infinitude of physical entities in a finite
region. To be sure, this particular setup is highly implausible; we
are bringing it into the discussion as an easy warmup case, before proceeding
to more physically plausible cases.

Now, what will happen as the projectile moves rightward?
Since there is nothing apparent to impede the projectile at negative positions
($x<0$), it would seem that it should continue its rightward motion until it
strikes a barrier. But before it can strike a barrier at $x=1/2^n$ it
must first strike (and pass through) all those to its left (at
$x=1/2^m$ for all $m>n$). This is impossible by the conditions of the
problem. So it cannot strike any barrier at all! Hence it must stop
its rightward motion, never passing zero, yet without touching
anything that would be a cause for its rightward motion to cease.

This has been debated in various philosophical papers; see
\cite{priest1999version, yablo2000reply, hansen2011new}. In \cite{perlis}
standard physics is brought to bear on the
puzzle in the forms of classical mechanics, quantum mechanics, and
relativity, showing for instance in the classical case that a
field effect in the form of a
repulsive force is mandated by Newton's Laws, so that the projectile
is bounced back to the left before passing zero. But the lesson for us
here is that even a dense infinity need not be paradoxical when seen
from within standard physical theory.  (Of course, one can resurrect a
paradox by insisting the barriers produce no forces outside their own
immediate locations; and the lesson then would be that this is
inconsistent with standard physics.)

Another version of the puzzle involves a continuous barrier-wall
extending from some point $b>0$ all the way back to, but not
including, $x=0$. That is, this is a wall of width $b$ but with its
left face missing. While a seeming bit of physical nonsense (at least in
terms of materials made of atoms) it is a familiar enough entity in
mathematics, essentially a half-open half-closed interval. And the
same form of argument applies as in the earlier Benardete example.
It would seem that physical entities cannot be isolated quite as well as our
imaginations might like: physical interactions will occur and cannot be
dismissed by mere stipulation.

Thus the Benardete examples provide a kind of dense infinity, but not
apparently one that ``breaks'' anything. Perhaps this is because it
does not directly involve an infinite density of standard physical
quantities like mass or charge or energy. (A closer analysis might
turn up an infinite sort of potential energy, however.) In any event,
when we turn to something ``real'' such as an electron, the situation
presents itself more starkly.

\section{The electron -- getting to the point}
An electron presents a somewhat related challenge. An electromagnetic
field exists around any charged particle. If the particle is not in
motion, then it is simply an electric field, $\vec{E}$, given by Coulomb's
Law. But the same law mandates that the field's magnitude $E$
increases at locations closer to the particle, approaching infinity in
the limit. In addition, the charge density is zero outside the
immediate location of the electron, and infinity at that
location. Finally, the mass density is also infinite at the location
of the electron, and zero elsewhere. These claims are based on the not uncommon
assumption that an electron has no spatial extent and is located at a
literal mathematical point; experimentally, the electron's radius is less than 
$10^{-22}$ meters \cite{meschede2007optics}.\footnote 
{But see for instance \cite{sasabe1992virtual}.}
A similar situation arises in the case of a black hole, where the mass
density becomes infinite at the mathematical point (singularity) of
the hole itself.\footnote
{See \cite{wiki:black} for an
  interesting discussion of electrons as black holes. A related set of
  issues involve the self-force and self-energy of an electron (or any
  point charge): the field created by a charge affects not only space
  surrounding the charge but also at the charge location(s) as
  well. Thus an electron's field influences it's own behavior. Similar
  considerations apply to any particle with non-zero mass: the
  associated gravitational field should affect the particle itself;
  see \cite{wald2011introduction}. }

One way to mathematically represent the situation of an infinite point density is via
a Dirac delta function, namely one that is infinite at the point in
question, and zero elsewhere. This -- usually taken as a convenient
fiction as already noted -- does the trick really well and
surprisingly often, and is now
a standard item in the physics toolbox. However, delta
functions can quickly turn from convenience to headache, due to the
nonlinearity of many applications. That is, the usual way to
``precisify'' a delta function is as a Schwartz distribution: a linear
functional on a space of functions. However -- as Wald \cite{wald2011introduction} points
out -- in many applications (nonlinear ones) delta functions (when
viewed as distributions) cannot be sensibly multiplied, and this
poses significant difficulties for their use where there are point
sources of fields.  This is a bit
outrageous: why cannot one multiply two functions? The answer is that
the Schwartz representation really groups these ``fiction-functions''
into equivalence classes (ones that provide the same results for
certain special integration properties\footnote
{Namely: $\int_{-\infty}^{+\infty}f_1(x)g(x)
  = \int_{-\infty}^{+\infty}f_2(x)g(x)$ for all ``test'' functions g.}), and, 
integration does not always respect some of the desired characteristics needed for non-linear applications.
Yet once ungrouped from each other and treated as genuine functions, delta functions can
indeed by multiplied, as we will see in the next section.\footnote
{This is not to say that successful application to non-linear differential equations is an automatic benefit; as noted, it is the product per se but rather integration properties of products that is at issue.}

Summarizing a bit, one way that infinity arises in physics is as
follows: a vector field (such as gravitational or electrostatic force)
depends on the spatial separation between one body and another, in a
way that increases without bound as that distance decreases to
zero. In particular, in these two instances, the force is proportional
to the reciprocal of the square of the distance. When that distance is
zero, the expression for the force becomes one divided by zero: $1/0$.

Now, division by zero is extremely problematic; it is not simply that
it is not defined, but that it is both overdetermined and
underdetermined. $0/0$ can be set equal to any number ($0/0=x$) with
impunity, since $0 = 0x$. And $1/0$ cannot be set equal to any number at
all, since $1 \neq 0x$. So there is no non-arbitrary nor even consistent
way to define division by zero that respects the basic concept of
division: $(a/b)b = a$, that is, as the inverse of multiplication.

It is tempting to say that this is because the real numbers are too
restrictive, and that  $1/0 = \infty$. But then what is $2/0$? And do we
allow $1 = 0 \times \infty$? These notions contain hints  of a possible
solution. In fact, mathematical physics often employs such intuitions,
in the form of infinitesimals and infinities; again think of the standard delta
function, that is zero at all non-zero reals, yet when infinitesimally
close to zero it rises up to infinity.

But mathematicians have invented many sorts of numbers, going well
beyond the familiar real and complex fields, including some that
explicitly contain infinities as first-class objects. Which fits the
physical situation best?  We shall not attempt to answer this here,
nor even to survey the existing options. Instead, we shall discuss
just one such option, with particular application to delta functions
and -- possibly -- to point particles.

\section{NSA}

One well-known approach to making sense of infinite and infinitesimal
quantities is nonstandard 
analysis (NSA), where the real number system {\bf R} is extended to
  {\bf *R}, which
includes ``numbers'' that are larger than every real, and also ones that
are smaller than every positive real and yet are themselves larger
than 0. The latter (small ones) and their negatives become the infinitesimals in common use in
physical reasoning. This was the aim of A. Robinson \cite{robinson1974non}: to develop
{\bf *R} and to show that in fact the familiar intuitive arguments using
infinitesimals then become quite rigorous.

But infinitesimals are not the same thing as zero; they are simply very very close to
zero; one might say that they form a kind of fuzzy zero -- and more
generally, that each real r has about it a band of new numbers (r plus
any infinitesimal) that ``coat'' r so closely that for ordinary
purposes r and its coat are indistinguishable.\footnote
{I apologize for introducing the term {\em coat} for this; already in
  use are: monad, haze, cloud, halo. My excuse is that a {\em coat of paint}
  is thin, hugs close to its target, and is not to be touched by other
  entities (at least while wet).}

A key point is that, while being in zero's coat, an infinitesimal
$\epsilon$  nonetheless has a
well-defined reciprocal $1/\epsilon$, which is infinite (larger than
every real). We still do not have a reciprocal for zero itself, but
perhaps we can dispense with that, and when a variable "approaches"
zero we may try to regard it as being in zero's coat rather than being zero
itself. More generally, the coat of a real r then provides stand-ins
for r,  which are r-ish in more or less degree (but all of them are
r-ish and not s-ish for any other real s).

As Robinson has shown, {\bf *R} can be given a very
rigorous definition, so that it remains an algebraic field and respects the
``usual'' mathematical properties of {\bf R}.  These  properties are given
sharp characterization, roughly as follows: for any sentence S that can be
expressed in a particular formal language L (including much of
standard math notations, for instance
+, $\times$, constants, =, $<$, $\forall$,  set-membership, etc -- but NOT using a symbol for
{\bf R} itself), S is true when interpreted as being about
elements in {\bf R} iff it is true about {\bf *R}.\footnote
{ Details can get a bit complicated; see \cite{davis2014applied}.} 
Now this
``transfer principle'' between {\bf R} and {\bf *R} is the basis for a
great many applications of NSA.\footnote
{There are by now dozens of books and hundreds or articles on the
  subject of NSA in general and   applications of the transfer
  principle in particular. See for  instance \cite{cutland2006nonstandard} and
  \cite{alberverio}.
}
But results of such applications -- at
least when those results are interpreted as being about {\bf R} (or
more precisely about the ``set-theoretic superstructure for {\bf R}'') --
generally are theorems that can also be proven (though maybe
less easily or intuitively) without NSA. One of the suggestions we are 
raising here is this: perhaps {\bf *R} (or its
superstructure) can be taken
seriously as a model of physical reality, to see whether this sheds
light on infinities that arise in physics.\footnote
{
See \cite{gudder1994toward} for a rare exceptional -- but alas all too preliminary -- treatment of NSA's nonstandard universe itself as having physical significance, in this case to QFT. 
}

One very nice (traditional) application of NSA is the delta function, which now can
be defined an as actual (non-fictional) function from {\bf *R} to {\bf
  *R}. For instance, given an infinitesimal $\epsilon$, let
$\delta(x) = 0$ for all numbers (in {\bf *R}) that lie outside $[-\epsilon/2,\epsilon/2]$,
and let $\delta(x)= 1/\epsilon$ for numbers in that interval.  The
graph of such a function
then is an infinitesimally thin, infinitely high rectangle, and the area
under it is exactly $\epsilon \times 1/\epsilon = 1$.  And then the integral of $\delta(x)$
times any function *f from {\bf *R} to {\bf *R} (that is an
appropriate extension of an integrable function f on the reals), gives f(0) -- or
more precisely, gives the
average value of *f in that interval, which is itself in the coat of --
and so normally indistinguishable from --  f(0).  

But now the {\em product} of any two such delta functions from {\bf *R} to {\bf
*R} is unproblematically another 
function from {\bf *R} to {\bf *R}. There is a tradeoff, however. For we
must {\em choose} a particular delta function to use in a given
application, rather than opt for the distributional approach that
lumps many such together.\footnote
{Further investigation (I am unaware of any work on this topic)  may
  reveal advantages to particular ``natural'' 
  choices for a delta function in particular applications. For now I
  simply point out one from Robinson's book (p. 138): $\frac{1}{\sqrt{\epsilon 
    \pi}} \exp(- \frac{x^2}{\epsilon})$. For real values of $\epsilon$ this is
  just an ordinary Gaussian, which arises quite
  naturally in many situations, and has very nice mathematical
  properties. Possibly in the nonstandard realm it will also play a
  helpful role. Note that this is not claimed to resolve issues about non-linear applications where integration properties of products arise.
}

\section{Back to the real world}

Now we return to physics, and in particular to the electron. We regard
it as being a point, or rather, we take its radius to be in the coat
of 0 (or whatever point it is centered on). That
is, we will postulate it to be a ball of infinitesimal radius.  In
particular, let some $\epsilon_e$ be that radius, and assume its mass
m is uniformly
distributed.\footnote
{ Note that this means the ball will be a proper subset of the coat,
  since coats have no boundary; if they did, then for instance
  $2\epsilon_e$ would be outside the coat, which makes no sense for it
  too is infinitesimal.}
Now we will attempt to characterize its spin ($\hbar / 2$)as a
physical angular momentum L of actual rotation, namely with an angular
frequency $\omega$ so that we get the usual classical formula:
$$L = \hbar/2 = I\omega = (2/5) m \epsilon_e^2 \omega$$
Since $\epsilon_e$ is infinitesimal then $\omega$ must be infinite, since
the LHS is finite.

The idea of treating spin as a possible rotational phenomenon was
considered long ago (see below), but taking the radius to be a
positive real r; this led to trouble with special relativity (SR). A
point on the surface of the electron ball would -- in order that the
rotation provide the proper angular momentum of spin, have to travel
at speeds in excess of the speed of light. But to reach such a speed
would require infinite energy, according to SR, and that traditionally
is taboo. Here then is a possible advantage of NSA: suppose we allow
physical quantities to be infinite.

Let's calculate the speed $v$ of a point on
the surface of an ``electron ball'' with (initially real) radius $r$
that is rotating with angular momentum $\hbar/2$. From the above equation, we get

\medskip
$v = \omega (1/2 \pi)(2 \pi r) = \omega r = 5\hbar / (4 m r)$
\medskip

If we insist that $v < c$ then we find 

\medskip
$c > 5\hbar/(4mr)$
\medskip

or

\medskip
$r > 5\hbar/(4mc) = 0.5 \times 10^{-12}$ meters.
\medskip

This is essentially the negative result found by Goudsmit and Uhlenbeck \cite{goudsmit}
that made them (and others) give up the idea of spin as deriving from
an actual physical rotation, since it was known even then that $r$ is
less than $3\times10^{-15}$ meters.\footnote
{But see for instance \cite{giulini2008electron} and \cite{ohanian1986spin}, for
  this is still a topic of dispute.} 

There is an alternative: allowing $v \geq c$, and also allowing infinite
energies, as well as replacing $r$ by $\epsilon_e$.  But why insist that $r$ be infinitesimal?
This is not strictly necessary. But since as already noted, it is
commonly thought that $r=0$ (an electron is an actual point with no
extent, no volume)\footnote
{For many purposes; but in QFT for instance this is not quite right.}
 and since we are allowing infinities anyway, it is
tempting to go ``all the way'' (at least all the way to infinitesimals,
if not literally to zero).

Back to our calculations: if $\epsilon_e$ is infinitesimal then as
noted above, the angular frequency $\omega$ is
infinite. But what then is the speed of a point in the electron coat,
at distance $\epsilon_e$ from the origin of rotation? It will be as
above, but replacing $r$ with $\epsilon_e$, hence infinite:
$$v = \omega \epsilon_e = 5\hbar / (4 m \epsilon_e) $$
This infinite speed precisely produces the finite angular momentum
$\hbar/2$. That is, the infinite speed  of a point within the
electron coat (which itself is at infinitesimal distance $\epsilon_e$ from the
origin of rotation), works together with that infinitesimal distance
to produce the needed finite angular momentum of spin.

However, not everything works out so nicely. 
The kinetic energy of mass m with speed $v$, in SR, is 
$$T = mc^2(\gamma - 1)$$
where $\gamma = 1 / \sqrt{1 - (v/c)^2)}$

When $v=c$, $\gamma$ is infinite, hence T would seem to be infinite. This is well-known,
of course, and is a primary reason that c is regarded as an
unreachable  upper limit on all speeds of massive objects. But it is
now even worse: for {\em this} infinity (of $\gamma$) seems to be of the totally
impossible kind: $1/0$.\footnote
{It is no good trying to wriggle out of this by
 supposing T is an NSA sort of infinity; that would correspond to v
 being ``almost'' the same as c (in the same coat, so that $v/c$ is in
 the coat of 1). For in fact we need -- for the Goudsmit/Uhlenbeck model
 -- that $v$ be even greater than c. And then $\gamma$ actually has an imaginary
 value! This leads into the even stranger physics of tachyons.}

There is however another interpretation: multiplying through by
$\sqrt{1 - (v/c)^2)}$, we get 
$$\sqrt{1 - (v/c)^2)} T = mc^2(1- \sqrt{1 - (v/c)^2)})$$
and for $v=c$ this becomes $0 \times T = mc^2$. A reasonable
conclusion now is that $m=0$: a particle traveling at light-speed has
no mass.\footnote
{This can actually be given a positive spin (pun intended). The Higgs
  field endows particles with mass according to whether they are
  retarded by it -- retarded from traveling at light-speed, that
  is. Particles that are not so retarded are by definition massless!}
And T is not further constrained here, infinite or
otherwise. Presumably it can (for $v=c$) be taken as the energy of an
appropriate light-speed particle.  Whether this is physical nonsense
or not, at least we are getting some sort of ``results'' from such an
approach.

\section{Summary and Discussion}


We have isolated three uses of infinities in physics: dense,
range-of-values, and useful-fiction. Range-of-values seems generally
unproblematic, but illustrative of the idea that our understanding of
the universe can involve infinities of some sort. These are not
directly measured, but rather are supported by a mix of inductive reasoning and
evidence; and they do not seem to present major difficulties.  

The infinities in the Benardete example perhaps lie in between
range-of-values and dense: many location values are posited, yet
they come close to representing an infinite density of something --
but it is not clear just what.  And there is no outright paradox if we
apply ordinary physics and a little commonsense.

But when we replace imagined barriers with actual physical entities
such as fields, things can quickly get bizarre, as in infinite values
for charge and force and mass densities. While our discussion focused
on a point-model of the electron, any point-source field will
do. There are standard tools for representing this -- for instance the
delta function -- but these are usually seen as merely useful
calculational devices and not as possible models of what the universe
is like. I am arguing that the great success of such tools speaks to
the strong possibility of an underlying phenomenon well-worth trying
to model.

I am not urging that NSA need be the mathematical physics of the
future. There are certainly other directions to consider, such as the
surreal numbers studied by Conway, Kruskal and others (see Wikipedia
entry \cite{wiki:surreal}). In addition, Bell \cite{bell1998primer} presents an approach
to infinitesimals (but not infinities) based on ``smooth worlds''
where logic (and geometry) gets even stranger than in NSA yet where
physics again comes into play.  And indeed infinity (of the dense
kind) might happen not to be physically sensible at all. But the idea
should not be discarded out of hand.


\bibliographystyle{plain}
\bibliography{tphis_arxiv}

\end{document}